\documentclass{article}
\usepackage{spconf,amsmath,amssymb,graphicx,booktabs}
\usepackage{microtype}
\AddToHook{cmd/thebibliography/after}{\setlength{\itemsep}{2pt}}
\usepackage[hidelinks]{hyperref}  
\usepackage{xcolor}
\usepackage{booktabs}
\usepackage{multirow}
\usepackage{graphicx}

\graphicspath{{figures/}}

\title{Automated Artifact Removal in EEG Age Prediction:\\A systematic comparison}

\name{\shortstack{\itshape
Davoud Hajhassani,
Paul-Adrien Graignic,
Bruno Aristimunha,
Apolline Mellot,
Tom Mariani,\\
\textit{Clément Nober,}
\textit{Bruna J. Lopes,}
\textit{Léo Burgund,}
\textit{Lionel Kusch,}
\textit{Thomas Semah,}
\textit{Arnault H. Caillet}
}}

\address{
Yneuro, Paris, France
}

\begin{document}
\begingroup
\setlength{\tabcolsep}{0pt}
\maketitle
\endgroup
\vspace{-8pt}
\begin{abstract}
\vspace{-8pt}
\noindent{Automated EEG artifact removal may improve downstream analysis but can also alter predictive information. We benchmarked nine automated artifact removal methods against a common no-artifact-removal baseline for cross-dataset EEG age prediction. We introduce Signal Quality Index (SQI)-guided GEDAI, which leverages local signal-quality assessment to restrict correction to the channel--epoch pairs requiring intervention. Three deep neural architectures were trained on TUEG and evaluated without target-domain fitting on ds005385, LEMON, and TDBRAIN. Across this setting, GEDAI and SQI-guided GEDAI were the only methods with consistent gains over baseline in age prediction performance across all datasets and architectures ($\Delta$MAE $=-0.77/-0.64$ years, $\Delta R^2=+0.083/+0.072$, respectively). The remaining methods were neutral or detrimental on average ($\Delta$MAE $=+0.22\pm0.16$ years, $\Delta R^2=-0.023\pm0.016$ across methods). The two GEDAI-based methods achieved closely matched performance, while SQI guidance reduced the median modification ratio from $74.78\%$ to $42.80\%$. These findings show that curation benefits are method-dependent and establish SQI guidance as a more selective operating point, leaving more of the original EEG unchanged and limiting the potential loss of neural activity while retaining most of GEDAI's predictive benefit.}
\end{abstract}


\vspace{-12pt}
\section{Introduction}
\label{sec:introduction}
\vspace{-8pt}
Electroencephalography (EEG) provides a non-invasive and relatively accessible window into brain function by capturing large-scale neuronal dynamics~\cite{buzsaki2012origin}. However, EEG is susceptible to physiological and instrumental artifacts, and variability in their handling can influence downstream analyses~\cite{mumtaz2021review}. As EEG datasets continue to grow, efficient, scalable, and reproducible automated artifact handling has become essential, yet selecting an appropriate method remains difficult~\cite{bomatter_machine_2024}. Existing automated artifact removal (AAR) methods differ substantially in how they identify, reject, or reconstruct contaminated data and in their effectiveness across artifact types~\cite{urigueen2015eeg}, motivating systematic comparison under common downstream conditions.

Artifact removal must balance artifact suppression with preservation of useful information~\cite{urigueen2015eeg}. Ocular, muscular, or cardiac activity may obscure neural signals, but can also act as relevant non-neural predictors~\cite{bomatter_machine_2024}. This is particularly relevant for deep learning (DL), which can exploit both neural and non-neural predictive structure and may therefore be sensitive to preprocessing choices.
Conversely, extensive EEG correction may attenuate genuine neural activity or remove task-relevant information, while rejection discards potentially usable EEG and limits online applicability~\cite{hajhassani2026improved}. Preserving as much unaltered EEG as possible is therefore an important objective.
Despite this, the effects of AAR on DL performance remain comparatively underexplored~\cite{roy2019deep}.

EEG age prediction provides a suitable downstream benchmark because chronological age is an objective continuous label shared across independent cohorts, enabling comparison of AAR methods across resting-state datasets without matched tasks, annotations or participant identities. Brain age prediction investigation is biologically meaningful, as deviations from expected age potentially inform on neurological health and disease severity~\cite{liem2027brainage}. Cross-dataset evaluation further tests whether predictive representations transfer across cohorts, recording systems, and acquisition conditions~\cite{engemann2022reusable,tveitstol2026robustness}.

We therefore benchmark nine AAR methods against a common no-artifact-removal baseline using three deep neural architectures and subsets of four public EEG datasets comprising 6,383 recordings from 2,971 healthy participants.
We further introduce SQI-guided GEDAI, a selective artifact-removal approach that applies correction exclusively to channel--epoch pairs identified as artifact-contaminated~\cite{hajhassani2026eeg}. We ask whether AAR consistently improves downstream prediction, and whether SQI guidance can retain the predictive benefit of correction methods while leaving more of the original EEG unchanged.
\vspace{-12pt}
\section{Methods}
\label{sec:method}
\vspace{-8pt}
\subsection{Datasets and Preprocessing}
\label{sec:Datasets and Preprocessing}
\vspace{-8pt}

\begin{figure}[t!]
    \centering
    \includegraphics[width=0.90\columnwidth,trim=0 2bp 0 3bp,clip]{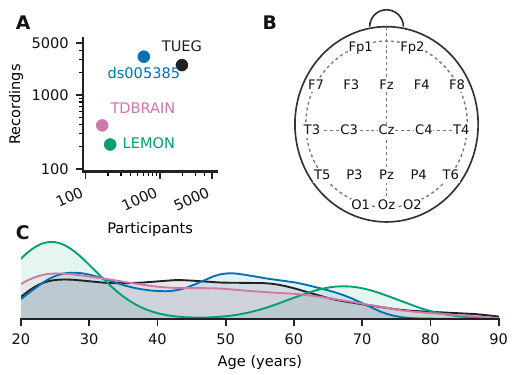}
    \caption{\small Dataset overview. (A) Number of participants and recordings per dataset. (B) Montage of the 20 selected EEG channels. (C) Age distribution by dataset. Colours identify datasets across panels.}
    \label{fig:datasets}
    \vspace{-10pt}
\end{figure}

We selected resting-state recordings of at least 60~s from healthy participants aged 20--90~years from four public EEG datasets: Temple University EEG Corpus (TUEG)~\cite{obeid2016tuh}, OpenNeuro ds005385~\cite{gajewski2022impact}, LEMON~\cite{babayan2019lemon}, and TDBRAIN~\cite{vandijk2022tdbrain}.
These datasets were selected for their relatively large sample sizes, broad age distributions, and previous use in EEG-based age prediction~\cite{engemann2022reusable}; their characteristics are shown in Fig.~\ref{fig:datasets}.

For each recording, up to the first 240~s were retained. Signals were band-pass filtered between 1 and 45~Hz, resampled to 200~Hz, average-referenced, and segmented into non-overlapping 5-s epochs. The selected AAR methods were then applied to correct artifact-contaminated signals or reject entire epochs, depending on the method. For the retained epochs, spatial sampling was harmonized by interpolating the signals onto 20 standard 10--05 channel locations (Fig.~\ref{fig:datasets}B) using spherical splines with $10^{-5}$ regularization~\cite{perrin1989spherical}. Finally, each channel was z-normalized across time within each epoch to standardize the inputs to the age prediction models.

\vspace{-8pt}
\subsection{Automated artifact removal methods}
\vspace{-8pt}

The AAR methods were selected to operate without manual artifact identification or expert intervention. ICLabel classifies components from independent component analysis and excludes artifactual components from signal reconstruction~\cite{piontonachini2019iclabel}. PREP identifies noisy channels based on unusual amplitudes, weak correlations, and poor predictability~\cite{bigdelyshamlo2015prep}. Autoreject uses cross-validation and Bayesian optimization to learn peak-to-peak thresholds for channel interpolation or epoch rejection~\cite{jas2017Autoreject}. Isolation Forest (IF) rejects outlying epochs without a manually defined amplitude threshold~\cite{zhang2024isolation}. FAAR is a lightweight method that combines artifact-sensitive features into an SQI and adaptively rejects contaminated epochs~\cite{hajhassani2026eeg}. ASR reconstructs high-variance subspaces relative to clean calibration data~\cite{chang2020asr}. RELAX combines amplitude- and gradient-based artifact detection with covariance-based multichannel Wiener filtering~\cite{bailey2023relax}. GEDAI identifies and removes artifact-related spatial components using generalized eigendecomposition with a leadfield-derived covariance reference, then reconstructs the clean EEG~\cite{ros2025gedai}. In our implementation, PREP channel rejection was disabled, Autoreject used interpolation counts $\{1,4,20\}$, and ASR used a cutoff of 10. Default parameters were used for all other settings.

\noindent\textbf{SQI-guided GEDAI.}
Because spatial decomposition may attenuate neural activity when artifact separation is imperfect, we extend GEDAI~\cite{ros2025gedai} with an SQI framework~\cite{hajhassani2026eeg} that restricts correction to channel--epoch pairs identified as artifact-contaminated. For each pair, SQI combines RMS amplitude, maximum temporal gradient, kurtosis, zero-crossing rate, band-limited spectral magnitude, robust standard deviation, and spectral robust noise-to-signal ratio. Descriptors are standardized against a recording-specific reference derived from automatically selected 1-s windows~\cite{hajhassani2026eeg,zhao2023quantitative}, mapped to artifact-severity levels, summed, and normalized to $[0,1]$, where higher values indicate poorer signal quality. Selective correction is applied as
$Y_{ec}(t)=
\begin{cases}
X_{ec}(t), & Q_{ec}<\tau\\
\widetilde{X}_{ec}(t), & Q_{ec}\geq\tau
\end{cases} ,
$ 
where $Q_{ec}$ is the SQI for channel $c$ and epoch $e$, and $X_{ec}(t)$ and $\widetilde{X}_{ec}(t)$ are the original and GEDAI-corrected signals. The threshold \mbox{$\tau=0.05$} was selected through grid search to minimize the validation mean absolute error (MAE). GEDAI correction is estimated from the complete multichannel epoch before SQI masking, so all channels contribute to artifact estimation while only channel--epoch pairs exceeding the SQI threshold are modified.

\begin{figure*}[t!]
    \centering
    \includegraphics[width=\textwidth,trim=0 1bp 0 9bp,clip]{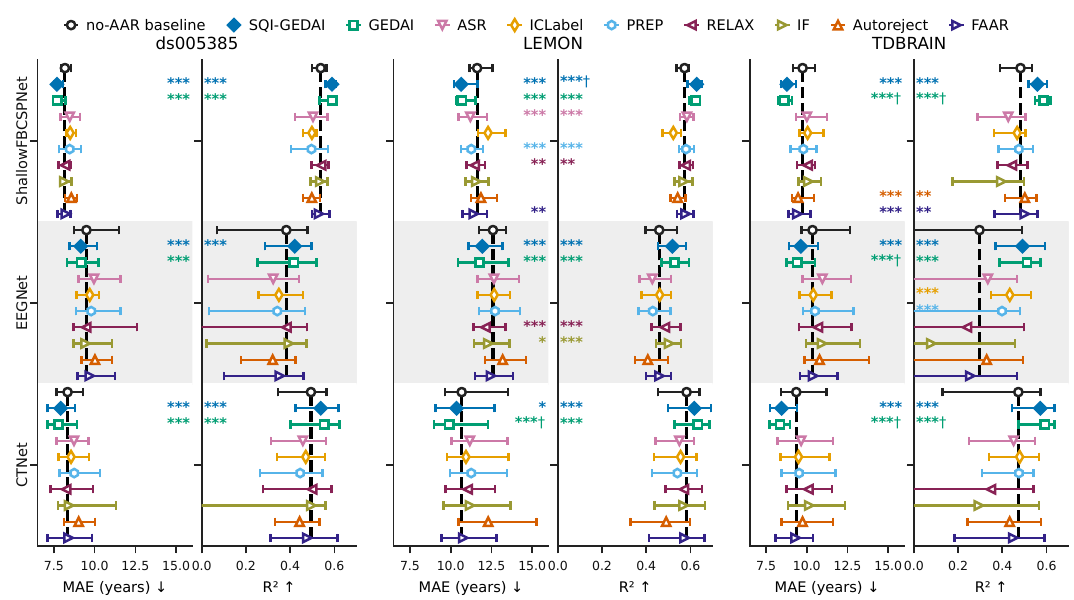}
    \caption{\small Participant-level age-regression performance across datasets and architectures. Markers and bars show the median and range across seeds, respectively; dashed lines indicate no-AAR medians. Stars denote improvement over no-AAR ($^{*}p_{\mathrm{Holm}}<0.05$, $^{**}p_{\mathrm{Holm}}<0.01$, $^{***}p_{\mathrm{Holm}}<0.001$), and $^{\dagger}$ denotes a difference between GEDAI and SQI-GEDAI ($p_{\mathrm{Holm}}<0.05$).}
    \label{fig:mae_r2}
    \vspace{-10pt}
\end{figure*}

\vspace{-8pt}
\subsection{Benchmark protocol and evaluation}
\vspace{-8pt}

\noindent\textbf{Cross-dataset protocol.}
The effects of AAR methods on age prediction were evaluated across datasets acquired with different devices and recording conditions. Models were trained on TUEG, the largest dataset by participant count, and evaluated without further tuning on ds005385, LEMON, and TDBRAIN. TUEG was split by participant into 80\% training and 20\% validation sets to prevent recording-level leakage. Each remaining dataset was used in full as an independent test set. Splitting and training were repeated over 50 random seeds to capture variability from data partitioning and stochastic training.

\noindent\textbf{Age-regression models.}
We evaluated ShallowFBCSPNet~\cite{schirrmeister2017deep}, EEGNet~\cite{lawhern2018eegnet}, and CTNet~\cite{zhao2024ctnet} using default Braindecode implementations~\cite{braindecode}, spanning spectral CNN, compact CNN, and convolutional-transformer families. Models were trained with $L_1$ loss and Adam, batch size 128, and a cosine learning-rate schedule peaking at $3.125\times10^{-4}$~\cite{gemein2024brain}. Training lasted up to 80 epochs with early stopping.

\noindent\textbf{Evaluation metrics.}
For each method, dataset, architecture, and seed, window-level predictions were averaged within recordings and then across recordings per participant, yielding one predicted age $\hat y_i$. We report $\mathrm{MAE}=N^{-1}\sum_i|y_i-\hat y_i|$ and the coefficient of determination $R^2=1-\sum_i(y_i-\hat y_i)^2/\sum_i(y_i-\bar y)^2$ over the $N$ participants, where $y_i$ is chronological age and $\bar y$ its mean. Upper-tail error was $P_{90}=Q_{0.90}(\{|y_i-\hat y_i|\})$, computed per seed from participant-level absolute errors. Paired differences, $\Delta P_{90}=P_{90}^{\mathrm{method}}-P_{90}^{\mathrm{baseline}}$, were averaged across seeds. Negative values indicate lower upper-tail error than no-AAR.


\noindent\textbf{Statistical analysis.}
Identical participant splits and random seeds across methods enabled paired comparisons. Improvements over no-AAR were tested using one-sided paired Wilcoxon signed-rank tests, with lower MAE or higher $R^2$ indicating improvement, while GEDAI and SQI-guided GEDAI were compared two-sided. Holm correction was applied within each metric and comparison family at $\alpha=0.05$. Overall mean paired differences and their 95\% confidence intervals (CIs) used 20{,}000 paired seed-ID bootstrap resamples for baseline and direct method comparisons, with target participants fixed and dataset--architecture combinations equally weighted.

\noindent\textbf{Modification ratio.}
The recording-level modification ratio measured the percentage of channel--epoch pairs altered immediately by each AAR method. A pair was modified if its epoch was removed or its maximum absolute change exceeded 1\% of the within-pair 95th percentile absolute amplitude: $\max_t|X_{ec}(t)-Y_{ec}(t)|>0.01\,P_{95}(|X_{ec}|)$. For Autoreject, the ratio includes both interpolated channel--epoch pairs and rejected epochs, counting each pair only once. For each method, we report the median ratio across all recordings.

For computational and environmental context, the complete experimental campaign was estimated to emit 91~kg~CO$_2$-eq using emission factors from RTE éCO$_2$mix.

\vspace{-12pt}
\section{Results}
\label{sec:results}
\vspace{-8pt}

\noindent\textbf{Age prediction performance.}
GEDAI and SQI-guided GEDAI ranked first or second for MAE and $R^2$ in all nine dataset--architecture combinations. MAE/$R^2$, averaged across cell-wise seed medians, were 9.27/0.559 and 9.40/0.548, respectively, versus 10.04/0.476 for no-AAR. Improvements over no-AAR were significant in 17/18 GEDAI and 18/18 SQI-GEDAI comparisons, versus only 8/63 MAE and 9/63 $R^2$ comparisons for the remaining methods (Fig.~\ref{fig:mae_r2}). Without target-domain fitting, the best-performing configuration per dataset achieved median-across-seed MAE/$R^2$ of 7.69/0.589 on ds005385 (SQI-guided GEDAI/ShallowFBCSPNet), 9.91/0.632 on LEMON (GEDAI/CTNet), and 8.36/0.592 on TDBRAIN (GEDAI/CTNet).

\noindent\textbf{Retraining variability.}
SQI-GEDAI/GEDAI showed lower across-seed variability than no-AAR in most dataset--architecture combinations (MAE: 7/9 and 6/9; $R^2$: 8/9 for both). Mean SDs were also lower (MAE: 0.36/0.37 vs.\ 0.42 years; $R^2$: 0.032/0.031 vs.\ 0.062; Fig.~\ref{fig:mae_r2}), whereas the remaining methods were generally more variable than no-AAR.

\noindent\textbf{Effect of SQI guidance.}
On average, SQI-guided GEDAI had a small 0.11-year higher MAE than GEDAI (95\% CI: $[0.06,0.15]$; cell-wise range: $[-0.08,0.35]$) and a 0.008 lower $R^2$ (95\% CI: $[-0.012,-0.003]$). Only 7 of 18 cell-wise paired Wilcoxon tests were significant after Holm correction: six favored GEDAI and one favored SQI-guided GEDAI (Fig.~\ref{fig:mae_r2}).

\begin{figure}[!t]
    \centering
    \includegraphics[width=\columnwidth,trim=0 2bp 0 6bp,clip]{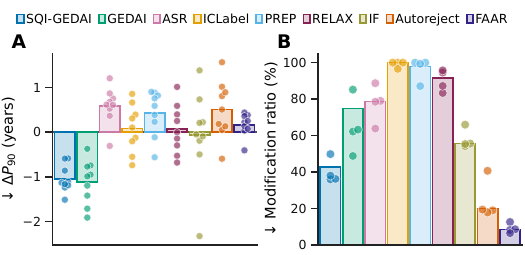}
    \caption{\small Upper-tail error and modification ratio. (A) Mean $\Delta P_{90}$ across seeds for each dataset--architecture combination; bars show the overall average. (B) Median modification ratios across all recordings; points show dataset-specific medians.}
    \label{fig:upper_tail_modif_ratio}
    \vspace{-10pt}
\end{figure}

\noindent\textbf{Upper-tail error.}
GEDAI and SQI-guided GEDAI were the only methods with negative $\Delta P_{90}$ across all nine dataset--architecture combinations, with mean reductions of $-1.12$ and $-1.05$~years, respectively (Fig.~\ref{fig:upper_tail_modif_ratio}A).

\begingroup
\widowpenalty=10000
\noindent\textbf{Modification ratio.}
Across all recordings, SQI-guided GEDAI had a median modification ratio of 42.80\%, compared with 74.78\% for GEDAI, a reduction of 31.98 percentage points. Its ratio was lower than those of all other AAR methods (55.56--100\%) except Autoreject and FAAR, which had median ratios of 19.97\% and 8.33\% respectively (Fig.~\ref{fig:upper_tail_modif_ratio}B).
\par
\endgroup

\vspace{-12pt}
\section{Discussion}
\label{sec:discussion}
\vspace{-8pt}

In this benchmark, we systematically investigated how automated EEG artifact removal affects cross-dataset age prediction. We also examined whether SQI guidance could maintain predictive performance while leaving more of the input EEG unchanged, aiming to limit the potential loss of neural activity.

\noindent\textbf{Artifact-removal benefits are method-dependent and require downstream validation.}
Artifact removal was not inherently beneficial for cross-dataset age prediction. GEDAI and SQI-guided GEDAI were the only methods that consistently improved median MAE and $R^2$ across datasets and architectures, whereas the other seven methods produced significant gains in only 17 of 126 baseline comparisons and were generally neutral or detrimental on average (Fig.~\ref{fig:mae_r2}). The two GEDAI-based methods also reduced upper-tail error across all nine dataset--architecture combinations, showing that their benefits extended beyond average accuracy to the upper-tail error threshold and the most poorly predicted participants (Fig.~\ref{fig:upper_tail_modif_ratio}A). Their gains across all three DL architectures suggest these benefits were not tied to a specific model family, although the magnitude of the gains varied across datasets. The lower across-seed variability of GEDAI-based methods also indicates greater robustness to source splits and stochastic training (Fig.~\ref{fig:mae_r2}). One plausible explanation is that GEDAI-based curation yields a more consistent input representation across retraining than other methods.

Because all methods used matched participant splits, random seeds, and target cohorts, these differences isolate the effect of AAR under a common protocol. Performance below no-AAR therefore suggests that curation removed or altered information useful for age prediction, reduced the available data, or introduced transformations that transferred poorly across acquisition conditions. However, downstream performance alone cannot determine whether the affected information was neural or artifactual, nor does underperformance imply that these methods are ineffective for their intended artifact removal objectives. Taken together, Automated EEG artifact removal should therefore be validated for the intended downstream task rather than assumed beneficial, as gains differed by method and varied in magnitude across datasets.


Our best GEDAI-based models reached $R^2=0.632$ on LEMON and $R^2=0.592$ on TDBRAIN. A previous cross-dataset study reported best single-source $R^2$ of $0.65$ and $0.48$ on these datasets, respectively~\cite{tveitstol2026robustness}. Our results were therefore close on LEMON and numerically higher on TDBRAIN, although differences in training data and evaluation protocols prevent a direct comparison.


\noindent\textbf{Selective correction reduces intervention while retaining epochs.}
SQI-guided GEDAI retained GEDAI's predictive benefit while reducing the median modification ratio from 74.78\% to 42.80\% (Fig.~\ref{fig:upper_tail_modif_ratio}B), showing that many corrections can be avoided without compromising prediction. This matters because unnecessary modification may distort task-relevant neural activity and should be avoided when possible~\cite{hajhassani2026eeg}. However, minimizing intervention alone is not sufficient: FAAR achieved the lowest modification ratio (8.33\%) by discarding contaminated epochs rather than correcting them, but did not match the predictive gains of the GEDAI-based methods and reduced the data available downstream. Low modification alone is insufficient. SQI-guided GEDAI combines three desirable properties by leaving acceptable EEG unchanged, retaining contaminated epochs through targeted correction, and achieving best predictive performance. These results support selective correction as a better balance than either broad correction or strict artifact rejection. Modification ratio should therefore be interpreted jointly with predictive performance and intervention type, since it measures how often EEG is altered rather than correction magnitude or neural preservation.

\noindent\textbf{SQI provides an actionable quality-control layer.}
In deployed EEG systems, the channel--epoch SQI can be used as a local decision variable, beyond its selective reconstruction role, to decide when to correct EEG, flag poor electrode quality, trust or withhold model outputs, and suspend updates to adaptive models or authentication templates. Separate thresholds can govern each decision independently, providing explicit quality-based control over downstream system behavior.

\noindent\textbf{Limitations and future directions.}
This study considered one source corpus, one downstream task, and a fixed SQI threshold selected on source validation data. The modification ratio measures how often EEG is altered, but not correction magnitude, artifact attenuation, or neural preservation. Future work should evaluate recording-adaptive thresholding, extend to other SQI-guided correction methods, and real-time feasibility. Artifact-annotated, controlled-contamination, or clean-reference recordings are also needed to determine whether selective correction better preserves genuine neural activity.

\noindent\textbf{Acknowledgements.} This work was supported by computing and storage resources provided by GENCI--IDRIS under allocation 2025-A0201016987 on the A100 partitions of the Jean Zay supercomputer.

\clearpage

\bibliographystyle{IEEEbib}
\bibliography{refs}

@article{obeid2016tuh,
  author  = {Obeid, Iyad and Picone, Joseph},
  title   = {The {T}emple {U}niversity {H}ospital {EEG} Data Corpus},
  journal = {Front. Neurosci.},
  volume  = {10},
  pages   = {196},
  year    = {2016},
  doi     = {10.3389/fnins.2016.00196}
}

@article{bomatter_machine_2024,
  author  = {Bomatter, Philipp and others},
  title   = {Machine learning of brain-specific biomarkers from {EEG}},
  journal = {eBioMedicine},
  volume  = {106},
  pages   = {105259},
  year    = {2024},
  doi     = {10.1016/j.ebiom.2024.105259}
}

@article{gajewski2022impact,
  author  = {Gajewski, Patrick and others},
  title   = {Impact of Biological and Lifestyle Factors on Cognitive Aging and Work Ability in the {D}ortmund {V}ital Study: Protocol of an Interdisciplinary, Cross-sectional, and Longitudinal Study},
  journal = {JMIR Res. Protoc.},
  volume  = {11},
  number  = {3},
  pages   = {e32352},
  year    = {2022},
  doi     = {10.2196/32352}
}

@article{babayan2019lemon,
  author  = {Babayan, Anahit and others},
  title   = {A mind-brain-body dataset of {MRI}, {EEG}, cognition, emotion, and peripheral physiology in young and old adults},
  journal = {Sci. Data},
  volume  = {6},
  pages   = {180308},
  year    = {2019},
  doi     = {10.1038/sdata.2018.308}
}

@article{vandijk2022tdbrain,
  author  = {van Dijk, Hanneke and others},
  title   = {The two decades brainclinics research archive for insights in neurophysiology {(TDBRAIN)} database},
  journal = {Sci. Data},
  volume  = {9},
  pages   = {333},
  year    = {2022},
  doi     = {10.1038/s41597-022-01409-z}
}

@article{piontonachini2019iclabel,
  author  = {Pion-Tonachini, Luca and others},
  title   = {{ICLabel}: An automated electroencephalographic independent component classifier, dataset, and website},
  journal = {NeuroImage},
  volume  = {198},
  pages   = {},
  year    = {2019},
  doi     = {10.1016/j.neuroimage.2019.05.026}
}

@article{ros2025gedai,
  author  = {Ros, Tomas and others},
  title   = {Return of the {GEDAI}: Unsupervised {EEG} Denoising Based on Leadfield Filtering},
  journal = {bioRxiv},
  year    = {2025}
}

@article{jas2017autoreject,
  author  = {Jas, Mainak and others},
  title   = {Autoreject: Automated artifact rejection for {MEG} and {EEG} data},
  journal = {NeuroImage},
  volume  = {159},
  pages   = {417--429},
  year    = {2017},
  doi     = {10.1016/j.neuroimage.2017.06.030}
}

@article{chang2020asr,
  author  = {Chang, Chih-Yun and others},
  title   = {Evaluation of Artifact Subspace Reconstruction for Automatic Artifact Components Removal in Multi-Channel {EEG} Recordings},
  journal = {IEEE Trans. Biomed. Eng.},
  volume  = {67},
  number  = {4},
  pages   = {1114--1121},
  year    = {2020},
  doi     = {10.1109/TBME.2019.2930186}
}

@article{bigdelyshamlo2015prep,
  author  = {Bigdely-Shamlo, Nima and others},
  title   = {The {PREP} pipeline: standardized preprocessing for large-scale {EEG} analysis},
  journal = {Front. Neuroinform.},
  volume  = {9},
  pages   = {16},
  year    = {2015},
  doi     = {10.3389/fninf.2015.00016}
}

@article{zhang2024isolation,
  author  = {Zhang, Runkai and others},
  title   = {Reliable and fast automatic artifact rejection of Long-Term {EEG} recordings based on Isolation Forest},
  journal = {Med. Biol. Eng. Comput.},
  volume  = {62},
  number  = {2},
  pages   = {521--535},
  year    = {2024},
  doi     = {10.1007/s11517-023-02961-5}
}

@article{bailey2023relax,
  author  = {Bailey, Neil W. and others},
  title   = {Introducing {RELAX}: An automated pre-processing pipeline for cleaning {EEG} data---Part 1: Algorithm and application to oscillations},
  journal = {Clin. Neurophysiol.},
  volume  = {149},
  pages   = {178--201},
  year    = {2023},
  doi     = {10.1016/j.clinph.2023.01.017}
}

@article{schirrmeister2017deep,
  author  = {Schirrmeister, Robin Tibor and others},
  title   = {Deep learning with convolutional neural networks for {EEG} decoding and visualization},
  journal = {Hum. Brain Mapp.},
  volume  = {38},
  number  = {11},
  pages   = {},
  year    = {2017},
  doi     = {10.1002/hbm.23730}
}

@article{lawhern2018eegnet,
  author  = {Lawhern, Vernon J. and others},
  title   = {{EEGNet}: a compact convolutional neural network for {EEG}-based brain--computer interfaces},
  journal = {J. Neural Eng.},
  volume  = {15},
  number  = {5},
  pages   = {},
  year    = {2018},
  doi     = {10.1088/1741-2552/aace8c}
}

@article{zhao2024ctnet,
  author  = {Zhao, Wei and others},
  title   = {{CTNet}: A convolutional transformer network for {EEG}-based motor imagery classification},
  journal = {Sci. Rep.},
  volume  = {14},
  pages   = {20237},
  year    = {2024},
  doi     = {10.1038/s41598-024-71118-7}
}

@article{hajhassani2026eeg,
  author  = {Hajhassani, Davoud and others},
  title   = {From {EEG} Cleaning to Decoding: The Role of Artifact Rejection in {MI}-based {BCI}s},
  journal = {arXiv},
  year    = {2026}
}

@article{zhao2023quantitative,
  author  = {Zhao, L. and others},
  title   = {Quantitative signal quality assessment for large-scale continuous scalp electroencephalography from a big data perspective},
  journal = {Physiol. Meas.},
  volume  = {44},
  number  = {3},
  pages   = {035009},
  year    = {2023},
  doi     = {10.1088/1361-6579/ac890d}
}

@article{roy2019deep,
  author  = {Roy, Yannick and others},
  title   = {Deep learning-based electroencephalography analysis: a systematic review},
  journal = {J. Neural Eng.},
  volume  = {16},
  number  = {5},
  pages   = {051001},
  year    = {2019},
  doi     = {10.1088/1741-2552/ab260c}
}

@article{urigueen2015eeg,
  author  = {Urig{\"u}en, Jose A. and Garcia-Zapirain, Bego{\~n}a},
  title   = {{EEG} artifact removal---state-of-the-art and guidelines},
  journal = {J. Neural Eng.},
  volume  = {12},
  number  = {3},
  pages   = {031001},
  year    = {2015},
  doi     = {10.1088/1741-2560/12/3/031001}
}

@article{buzsaki2012origin,
  author  = {Buzs{\'a}ki, Gy{\"o}rgy and others},
  title   = {The origin of extracellular fields and currents---{EEG}, {ECoG}, {LFP} and spikes},
  journal = {Nat. Rev. Neurosci.},
  volume  = {13},
  number  = {6},
  pages   = {407--420},
  year    = {2012},
  doi     = {10.1038/nrn3241}
}

@article{hajhassani2026improved,
  author  = {Hajhassani, Davoud and others},
  title   = {Improved {R}iemannian potato field: An automatic artifact rejection method for {EEG}},
  journal = {Biomed. Signal Process. Control},
  year    = {2026}
}

@article{engemann2022reusable,
  author  = {Engemann, Denis A. and others},
  title   = {A reusable benchmark of brain-age prediction from {M/EEG} resting-state signals},
  journal = {NeuroImage},
  volume  = {262},
  pages   = {119521},
  year    = {2022},
  doi     = {10.1016/j.neuroimage.2022.119521}
}

@article{perrin1989spherical,
  author  = {Perrin, Fran{\c{c}}ois M. and others},
  title   = {Spherical splines for scalp potential and current density mapping},
  journal = {Electroencephalogr. Clin. Neurophysiol.},
  volume  = {72},
  number  = {2},
  pages   = {184--187},
  year    = {1989},
  doi     = {10.1016/0013-4694(89)90180-6}
}

@misc{braindecode,
  author = {Aristimunha, Bruno and others},
  title = {Braindecode: toolbox for decoding raw electrophysiological brain data
           with deep learning models},
  howpublished = {Zenodo},
  url = {https://github.com/braindecode/braindecode},
  doi = {10.5281/zenodo.17699192},
  license = {BSD-3-Clause},
}

@article{mumtaz2021review,
  author  = {Mumtaz, Wajid and others},
  title   = {Review of challenges associated with the {EEG} artifact removal methods},
  journal = {Biomed. Signal Process. Control},
  volume  = {68},
  pages   = {102741},
  year    = {2021},
  doi     = {10.1016/j.bspc.2021.102741}
}

@article{tveitstol2026robustness,
  author  = {Tveitst{\o}l, Thomas and others},
  title   = {Assessing the Robustness of Deep Learning Based Brain Age Prediction Models Across Multiple {EEG} Datasets},
  journal = {IEEE Trans. on Bio. Eng.},
  year    = {2026},
  volume  = {73},
  number  = {7},
  pages   = {2582--2601},
  doi     = {10.1109/TBME.2025.3639477},
  url     = {https://doi.org/10.1109/TBME.2025.3639477}
}

@article{gemein2024brain,
  author  = {Gemein, Lukas A. W. and others},
  title   = {Brain age revisited: Investigating the state vs. trait hypotheses of {EEG}-derived brain-age dynamics with deep learning},
  journal = {Imaging Neurosci.},
  volume  = {2},
  pages   = {imag--2--00210},
  year    = {2024},
  doi     = {10.1162/imag_a_00210}
}

@article{liem2027brainage,
  author  = {Liem, Franziskus and others},
  title   = {Predicting brain-age from multimodal imaging data captures cognitive impairment},
  journal = {NeuroImage},
  volume  = {148},
  pages   = {179-188},
  year    = {2017},
  doi     = {10.1016/j.neuroimage.2016.11.005}
}

\end{document}